\documentclass[reprint]{revtex4-1}
\usepackage{mathtools}
\usepackage{amsmath}
\usepackage{amsfonts}
\usepackage{amssymb}
\usepackage{graphicx}
\usepackage{hyperref}
\usepackage{bm}

\begin{document}
\title{An apparent paradox concerning the field of an ideal dipole}
\author{Edward Parker}
\email{tparker@physics.ucsb.edu}
\affiliation{Department of Physics, University of California, Santa Barbara, CA 93106}
\date{\today}

\begin{abstract}
The electric or magnetic field of an ideal dipole is known to have a Dirac delta function at the origin.  The usual textbook derivation of this delta function is rather ad hoc and cannot be used to calculate the delta-function structure for higher multipole moments.  Moreover, a naive application of Gauss's law to the ideal dipole field appears to give an incorrect expression for the dipole's effective charge density.  We derive a general result for the delta-function structure at the origin of an arbitrary ideal multipole field without using any advanced techniques from distribution theory.  We find that the divergence of a singular vector field can contain a \emph{derivative} of a Dirac delta function even if the field itself does not contain a delta function.  We also argue that a physical interpretation of the delta function in the dipole field previously given in the literature is perhaps misleading and may require clarification.
\end{abstract}
\maketitle

\section{Introduction \label{Intro}}

There is a well-known subtlety in classical electromagnetism regarding the fields produced by point particles and other charge distributions localized at a single point.  Because the electric potential $\phi(\bm{x})$, electric field $\bm{E}(\bm{x})$, and charge density $\rho(\bm{x})$ typically diverge or are otherwise singular at the charge's location, an attempt to use naive differentiation to satisfy the identities $\bm{E} = -\bm{\nabla} \phi$ and $\bm{\nabla} \cdot \bm{E} = 4 \pi \rho$ (in CGS units), and the corresponding magnetostatic equations, often ``misses'' Dirac delta-function contributions to the derivatives.  The best-known example of this phenomenon involves the divergence of the electric field of a point charge: if we consider a point charge $q$ at the coordinate origin and define $r := |\bm{x}|$ and the radial unit vector $\bm{n} := \bm{x}/r$, then $\bm{\nabla} \cdot \bm{E}(\bm{x}) = \bm{\nabla} \cdot \left( q\, \bm{n} / r^2 \right)$ naively appears to vanish.  But in fact, a more careful treatment \cite{Jackson} of the divergence of $\bm{E}(\bm{x})$ at the origin shows that it actually contains a Dirac delta function $4 \pi q\, \delta^3(\bm{x})$, and so Gauss's law gives that the point particle's effective charge density is $\rho(\bm{x}) = q\, \delta^3(\bm{x})$.

Similar delta-function subtleties arise in the case of higher ideal multipoles, and unfortunately it becomes much more difficult to find mutually compatible expressions for $\phi$, $\bm{E}$, and $\rho$ (or the corresponding magnetic quantities).  In this article, we will examine in detail the simplest case beyond that of an ideal monopole - an ideal dipole - and find that even in this case, applying Gauss's law to find the effective charge density is not at all trivial.  In the rest of this section, we will give several arguments justifying the standard expression for the delta-function structure in the electric field of an ideal dipole, and give a physical consequence of this delta function.  In Section~\ref{Paradox}, we argue that two different methods for calculating the dipole's charge density $\rho(\bm{x})$ appear to give different results.  In Section~\ref{Discussion}, we generalize to the case of an arbitrary multipole and show that the paradox can be resolved by appropriately taking into account the \emph{far-field} contribution to the charge density at the origin.  In Section~\ref{Conclusion} we conclude.

If we expand the electric potential due to a localized source of electric charge in terms of multipole moments, we find that the dipole term is \cite{Jackson}
\begin{equation} \label{dipolePhi}
\phi(\bm{x}) = \frac{\bm{p} \cdot \bm{x}}{r^3} = \frac{p_i x_i}{r^3},
\end{equation}
where $\bm{p}$ is the charge distribution's dipole moment (chosen to lie at the coordinate origin) and repeated indices are summed.  Far away from the charge distribution, the far field due to the dipole moment is
\begin{equation} \label{ff}
E_{\text{ff},i}(\bm{x}) = \frac{3 n_i\, p_j n_j - p_i}{r^3}.
\end{equation}
For example, if the charge distribution consists of two particles with charge $q$ and $-q$ separated by a displacement vector $\bm{d}$ with $\bm{p} = q \bm{d}$, then $\bm{E}_\text{ff}$ describes the field at distances $r \gg d$.  In the idealized limit where $q \rightarrow \infty$ and $\bm{d} \rightarrow \bm{0}$ with their product $\bm{p}$ held constant, the potential is given exactly by \eqref{dipolePhi} for all $\bm{x} \neq \bm{0}$.

The fact that $\phi(\bm{x})$ diverges at the origin suggests that for an ideal point dipole, the far-field expression \eqref{ff} may need to be modified there.  Indeed, the correct expression is
\begin{equation} \label{E}
E_i(\bm{x}) = \frac{3 n_i\, p_j n_j - p_i}{r^3} - \frac{4 \pi}{3} p_i\, \delta^3(\bm{x}).
\end{equation}

One standard argument \cite{Jackson} justifying the delta-function term is that if $V$ is the interior of a sphere containing all of the electric charge in a system with net electric dipole moment $\bm{p}$, then
\begin{equation} \label{intE}
\int_V d^3x\, \bm{E}(\bm{x}) = -\frac{4 \pi}{3} \bm{p}.
\end{equation}
Evaluating the volume integral over \eqref{ff} is tricky, because the integrand diverges at the origin, so we must specify a regularization procedure.  In this case, the regularization procedure is just a precise specification of the order of limits in which the three improper integrals over the spatial coordinates are to be evaluated.  Different regularization procedures are useful in different contexts, but in this article we will only consider the simplest one, known as ``spherical regularization:'' the convention that all integrals near singularities are to be evaluated in spherical coordinates centered at the singularity, with the angular integrals performed first \cite{Hnizdo}.  (This is a higher-dimensional analog of the convention that all integrals over singular functions are given by their Cauchy principal values.  In the more familiar case of a monopole, the electric field only diverges as $1/r^2$ so the improper integral converges in spherical coordinates, and we do not need to specify a regularization procedure.)  We will assume (without loss of generality) for the rest of this article that the dipole is aligned parallel to the $z$-axis.  Then by symmetry, only the $z$-component of the integral in \eqref{intE} could be nonzero, and
\begin{align*}
& \hspace{10pt} \int_V d^3x \left( \frac{3 n_z\, p_j n_j - p_z}{r^3} \right) \\
&= p \int_V d^3x \left( \frac{3 \cos^2 \theta - 1}{r^3} \right) \\
&= 4 \pi p \int_0^R \frac{dr}{r} \int_0^\pi d\theta \sin \theta (3 \cos^2 \theta - 1) \\
&= 0,
\end{align*}
so the far-field term does not contribute to the integral in identity \eqref{intE}, and with this choice of regularization the expression \eqref{E} satisfies the identity.

A second argument \cite{Frahm} uses the identity $\partial_j (1/r) = -x_j/r^3$ to get
\begin{equation} \label{mixedPartialE}
E_i(\bm{x}) = -\partial_i\, \phi(\bm{x}) = -p_j \partial_i \left( \frac{x_j}{r^3} \right) = p_j \partial_i \partial_j \left( \frac{1}{r} \right).
\end{equation}
The mixed partial derivative is given by the identity \cite{Frahm, Bowen}
\begin{equation} \label{mixedPartials}
\partial_i \partial_j \left( \frac{1}{r} \right) = \frac{3 n_i n_j - \delta_{ij}}{r^3} - \frac{4 \pi}{3} \delta_{ij} \delta^3(\bm{x}),
\end{equation}
and \eqref{mixedPartialE} and \eqref{mixedPartials} together give \eqref{E}.  This identity was also explicitly derived using the spherical regularization procedure; different regularizations of the improper integral give expressions different from \eqref{mixedPartials} \cite{Hnizdo}.  (Roughly speaking, identity \eqref{intE} requires that the two terms in \eqref{E} must have a ``total'' delta function of $-(4 \pi / 3)\, \bm{p}\, \delta^3(\bm{x})$ between them, and the spherical regularization procedure puts it entirely into the second term.  Under a regularization procedure in which $\int_V d^3x\, \bm{E}_\text{ff}(\bm{x}) = c\, \bm{p}$, the coefficient of the delta function in \eqref{mixedPartials} becomes $-4 \pi / 3 - c$.)

A third argument \cite{Estrada, Namias, Blinder} modifies the potential \eqref{dipolePhi} by formally multiplying it by the unit step function $\theta(r)$ and using the fact that $d\theta / dr = \delta(r)$.  When done carefully, this technique generalizes the usual partial derivative to the ``distributional derivative'' from mathematical distribution theory.

The first argument is quite ad hoc - we simply noticed that the naive expression \eqref{ff} fails to satisfy one particular identity, and manually added a term to satisfy it.  The second argument is much more satisfying, as it allows the delta-function term and the far-field term to be derived simultaneously, but the identities corresponding to \eqref{mixedPartials} for higher multipole terms become increasingly complicated to calculate.  Ref.~\onlinecite{Estrada} presents the third argument rigorously, but requires the rather heavy-duty mathematical machinery of the distributional derivative.  Refs.~\onlinecite{Namias} and \onlinecite{Blinder} present it less rigorously, but use expressions like $(\bm{n} / r)\, \delta^3(\bm{x})$ and integrals in which a delta function lies exactly at one limit of integration, which arguably need to be treated more carefully.  Moreover, the techniques used in the third argument are also complicated to generalize to higher multipoles.  In Section~\ref{Discussion}, we present a single simple, intuitive calculation that gives the delta-function structure at the origin of an arbitrary multipole far field.

As physical motivation, the delta-function term in \eqref{E} has measureable effects.  The simplest one actually occurs in the corresponding expression for an ideal magnetic dipole.  The magnetic field $\bm{B}(\bm{x})$ is the curl of a vector potential, so if a sphere $V$ contains all the current in a system then $\int_V d^3x\, \bm{B}(\bm{x}) = (8 \pi / 3)\, \bm{m}$, where $\bm{m}$ is the current distribution's magnetic dipole moment.  We must therefore add to the magnetic dipole far-field term $\bm{B}_\text{ff}(\bm{x}) \equiv \bm{E}_\text{ff}(\bm{x})$ (with $\bm{p}$ replaced by $\bm{m}$) a term $(8 \pi / 3)\, \bm{m}\, \delta^3(\bm{x})$ under a spherical regularization procedure \cite{Jackson}.  In the nonrelativistic limit, particles with quantum-mechanical spin correspond to (so far as we know) ideal magnetic dipoles.  Moreover, a particle's wavefunction can probe the magnetic field precisely at another particle's location, so the delta-function term can affect the particles' interaction.  This can be seen most simply in the hyperfine splitting of the ground-state energy levels of the hydrogen atom due to the coupling between the proton's and electron's spins.  If we treat the spins' dipole-dipole interaction as a perturbation to the usual classical Coulomb potential, then it is straightforward to calculate that the first-order contribution to the hyperfine splitting is \cite{Griffiths}
\[
\Delta E_\text{hf} = \frac{8 \pi}{3} \frac{\gamma_e \gamma_p \hbar^2}{\pi a^3} = 5.884 \times 10^{-6} \text{ eV},
\]
where $\gamma_e$ and $\gamma_p$ represent the electron's and proton's gyromagnetic ratios, respectively, and $a$ is the Bohr radius. The prefactor $8 \pi / 3$ comes from the prefactor of the delta-function term in the ideal magnetic dipole field.  This energy level splitting is responsible for the famous 21-cm hydrogen line measured by radio astronomy, which is one of the most common forms of radiation in the universe and has been measured extremely accurately.  The prediction above agrees with experiment to 99.8\% accuracy, and quantum electrodynamics corrections further improve the accuracy \cite{Brodsky}.

For simplicity, we will now only consider electric multipole fields.  Similar considerations apply to magnetic multipoles, but the vector nature of the potential introduces mathematical complications that do not significantly affect our conclusions.

\section{An apparent paradox \label{Paradox}}

Another motivation for considering the delta-function structure at the origin more carefully and generally is given by an apparent paradox that arises in computing the effective charge density of an ideal dipole, which is
\begin{equation} \label{rho}
\rho(\bm{x}) = -\bm{p} \cdot \bm{\nabla} \delta^3(\bm{x}) = -p\, \delta(x) \delta(y) \delta'(z).
\end{equation}
The form of the expression is intuitively clear when we consider the ideal dipole as the limit of a physical dipole as $\bm{d} \rightarrow 0$ and $q \rightarrow \infty$, and that the distribution $\delta'(z)$ corresponds to a function that is strongly peaked at $(-\epsilon, p / \epsilon)$ and $(\epsilon, -p / \epsilon)$ for infinitesimal $\epsilon$.  We can derive it more rigorously in two different ways \cite{Jackson}.  One way is to note that
\begin{align*}
\phi(\bm{x}) &= \int d^3 x' \frac{\rho(\bm{x}')}{|\bm{x} - \bm{x}'|} = -\bm{p} \cdot \int d^3 x' \frac{\bm{\nabla}' \delta^3(\bm{x}')}{|\bm{x} - \bm{x}'|} \\
&= \bm{p} \cdot \int d^3 x'\, \delta^3(\bm{x}') \bm{\nabla}' \left( \frac{1}{|\bm{x} - \bm{x}'|} \right) =  \frac{\bm{p} \cdot \bm{x}}{r^3}
\end{align*}
in accordance with \eqref{dipolePhi} (where $\bm{\nabla}'$ denotes the gradient with respect to $\bm{x}'$).  A second, similar method is to verify that the potential energy of the dipole in an external potential $\phi_\text{ext}(\bm{x})$ (which does not include the potential from the dipole itself) gives the correct expression
\begin{align}
U &= \int d^3x\, \phi_\text{ext}(\bm{x})\, \rho(\bm{x}) = -\bm{p} \cdot \int d^3x\, \phi_\text{ext}(\bm{x}) \bm{\nabla} \delta^3(\bm{x}) \nonumber \\
&= \bm{p} \cdot \int d^3x\, \delta^3(\bm{x}) \bm{\nabla} \phi_\text{ext}(\bm{x}) = -\bm{p} \cdot \bm{E}_\text{ext}(\bm{0}). \label{U}
\end{align}

But applying Gauss's law to \eqref{E} gives
\begin{equation} \label{DelDotE}
\bm{\nabla} \cdot \bm{E}(\bm{x}) = \bm{\nabla} \cdot \bm{E}_\text{ff}(\bm{x}) - \frac{4 \pi}{3} \bm{p} \cdot \bm{\nabla} \delta^3(\bm{x}).
\end{equation}
The divergence $\bm{\nabla} \cdot \bm{E}_\text{ff}$ is clearly zero away from the origin because there is no charge away from the dipole.  $\bm{E}_\text{ff}(\bm{x})$ diverges at the origin, so we must specify a regularization procedure in order to evaluate its divergence there.  In order to be consistent with the derivations in Section~\ref{Intro}, we must again adopt the spherical regularization procedure in which integrals are performed in spherical coordinates and the angular integrals are performed first.  Under this procedure, $\bm{E}_\text{ff}$ does not ``contain'' a delta function at the origin (as discussed above), so taking the divergence of $\bm{E}_\text{ff}$ should not produce a derivative of a delta function.  We therefore seem to have that $\bm{\nabla} \cdot \bm{E}(\bm{x}) \overset{?}{=} -(4 \pi / 3)\, \bm{p} \cdot \bm{\nabla} \delta^3(\bm{x})$, from which Gauss's law $\bm{\nabla} \cdot \bm{E}(\bm{x}) = 4 \pi \rho(\bm{x})$ implies that $\rho(\bm{x}) \overset{?}{=} -(1 / 3)\, \bm{p} \cdot \bm{\nabla} \delta^3(\bm{x})$, which disagrees with \eqref{rho}.

This apparent paradox occurs under any choice of regularization, because as discussed in Section~\ref{Intro}, identity \eqref{intE} requires that the ``total'' delta function across both terms in \eqref{E} must be $-(4 \pi / 3)\, \bm{p}\, \delta^3(\bm{x})$, so the ``total'' derivative of a delta function in $\bm{\nabla} \cdot \bm{E}(\bm{x})$ should be $-(4 \pi / 3)\, \bm{p}\cdot \bm{\nabla} \delta^3(\bm{x})$, implying that $\rho(\bm{x}) \overset{?}{=} -(1 / 3) \bm{p} \cdot \bm{\nabla} \delta^3(\bm{x})$ under any choice of regularization.  We also cannot resolve the paradox by changing the prefactor of the delta-function term in \eqref{E} to $-4 \pi$, because doing so would contradict both \eqref{intE} and experimental results, as discussed above.

\section{Discussion \label{Discussion}}

Let us generalize to an arbitrary ideal multipole potential $\phi^{(lm)}(\bm{x})$ corresponding to a multipole moment $q_{lm}$, which we will define by
\[
\phi^{(lm)}(\bm{x}) := q_{lm} \frac{Y_{lm}(\Omega)}{r^{l+1}},
\]
where $\Omega$ denotes the angular coordinates, $Y_{lm}(\Omega)$ the usual spherical harmonics, and there is no sum on repeated indices.  (Note that we use a different normalization convention from Ref.~\onlinecite{Jackson} for $q_{lm}$.)  Away from the origin, the electric field is given by the far-field expression $\bm{E}^{(lm)}_\text{ff}(\bm{x}) = -\bm{\nabla} \phi^{(lm)}(\bm{x})$.

Delta functions and their derivatives are defined by their integrals against arbitrary smooth test functions $f(\bm{x})$, so in order to calculate $\bm{\nabla} \cdot \bm{E}_\text{ff}^{(lm)}(\bm{x})$ at the origin we must evaluate
\[
\int_V d^3x\, \left[ f(\bm{x}) \bm{\nabla} \cdot \bm{E}_\text{ff}^{(lm)}(\bm{x}) \right]
\]
over a volume $V$ that includes the origin.  We integrate by parts and supress the superscripts, subscripts, and arguments $\bm{x}$ for clarity:
\begin{equation} \label{IBP}
\int_V d^3x\, [f\, \bm{\nabla} \cdot \bm{E}] = \oint_{\partial V} d\bm{S} \cdot \left( f\, \bm{E} \right) - \int_V d^3x\, [\bm{E} \cdot \bm{\nabla} f].
\end{equation}
Only the neighborhood of the origin contributes to the integral on the LHS, so its value does not depend on the volume $V$ (as long as it contains the origin).  WLOG, we take $V$ to be a ball of radius $R$ centered at the origin.  We are interested in fields $\bm{E}$ that diverge at the origin, so we also need to specify a regularization procedure for the volume integral on the RHS of \eqref{IBP}.  We use spherical regularization again:
\begin{align}
\int_V d^3x\, [f\, \bm{\nabla} \cdot \bm{E}] &= R^2 \oint d\bm{\Omega} \cdot \left( f\, \bm{E} \right) \big|_{r = R} \label{intfDivF} \\
&\hspace{10pt} - \lim_{\epsilon \rightarrow 0^+} \int_\epsilon^R dr\, r^2 \oint d\Omega\, (\bm{E} \cdot \bm{\nabla} f). \nonumber
\end{align}

We now expand $f(\bm{x})$ in complex conjugate spherical harmonics:
\begin{equation} \label{f}
f(\bm{x}) = \sum_{l' = 0}^\infty \sum_{m' = -l'}^{l'} c_{l'm'}(r)\, Y^*_{l'm'}(\Omega).
\end{equation}
The polar angle dependence of $Y_{l'm'}^*(\Omega)$ is given by an order-$l'$ associated Legendre polynomial of $\sin \theta$ and $\cos \theta$, so in order for $f(\bm{x})$ to be smooth at the origin, $c_{l'm'}(r)$ must have a zero of order at least $l'$ at $r = 0$.

The radial component of the electric field is
\begin{equation} \label{Er}
E_r = (l + 1) \frac{q_{lm}}{r^{l+2}} Y_{lm}(\Omega),
\end{equation}
and because of the orthonormality of the spherical harmonics, the angular integration over the surface term in \eqref{intfDivF} gives
\begin{equation} \label{surf}
R^2 \oint d\bm{\Omega} \cdot \left( f\, \bm{E} \right) \big|_{r = R} = (l + 1)q_{lm} \frac{c_{lm}(R)}{R^l}.
\end{equation}

The volume term can be expressed as
\[
\int_\epsilon^R dr\, r^2 \oint d\Omega\, [(E_r \partial_r + \bm{E}_\perp \cdot \bm{\nabla}_\perp) f]
\]
where the subscript $\perp$ denotes the angular coordinates.  \eqref{f} and \eqref{Er} together give
\begin{equation} \label{radialTermAngularInt}
\oint d\Omega\, E_r \partial_r f = (l + 1) q_{lm} \frac{c'_{lm}(r)}{r^{l+2}}.
\end{equation}
We can also integrate
\begin{align*}
\oint d\Omega\, [(\bm{E}_\perp \cdot \bm{\nabla}_\perp) f] &= -\oint d\Omega\, \left[ \bm{\nabla}_\perp \phi^{(lm)} \cdot \bm{\nabla}_\perp f \right] \\
&= \oint d\Omega\, \left[ \phi^{(lm)} \nabla_\perp^2 f \right]
\end{align*}
by parts with no surface term because the surface of integration is closed.  Using the eigenvalue identity $\nabla_\perp^2 Y_{lm}(\Omega) = -(l(l+1) / r^2)\, Y_{lm}(\Omega)$, the angular integral gives
\begin{equation} \label{angularTermAngularInt}
\oint d\Omega\, [(\bm{E}_\perp \cdot \bm{\nabla}_\perp) f] = -l (l+1) q_{lm} \frac{c_{lm}(r)}{r^{l+3}}.
\end{equation}

Combining \eqref{radialTermAngularInt} and \eqref{angularTermAngularInt}, the volume term in \eqref{intfDivF} becomes
\begin{align}
&\hspace{11pt} \int_\epsilon^R dr\, r^2 \oint d\Omega\, (\bm{E} \cdot \bm{\nabla} f) \nonumber \\
&= (l + 1)\, q_{lm} \int_\epsilon^R dr\,\left[ \frac{c'_{lm}(r)}{r^l} - l \frac{c_{lm}(r)}{r^{l+1}} \right] \nonumber \\
&= (l + 1)\, q_{lm} \int_\epsilon^R dr\, \frac{d}{dr} \left( \frac{c_{lm}(r)}{r^l} \right) \label{volumeInt} \\
&= (l + 1)\, q_{lm} \left( \frac{c_{lm}(R)}{R^l} - \frac{c_{lm}(\epsilon)}{\epsilon^l} \right). \nonumber
\end{align}
The first term is cancelled by the surface term \eqref{surf}, so
\[
\int_V d^3x\, \left[ f\, \bm{\nabla} \cdot \bm{E}_\text{ff}^{(lm)} \right] = (l + 1)\, q_{lm} \lim_{\epsilon \rightarrow 0^+} \frac{c_{lm}(\epsilon)}{\epsilon^l}.
\]
Since $c_{lm}(r)$ has a zero of order at least $l$ at $r = 0$, this limit converges and we finally arrive at
\begin{equation} \label{master}
\int_V d^3x\, \left[ f\, \bm{\nabla} \cdot \bm{E}_\text{ff}^{(lm)} \right] = \frac{l + 1}{l!}\, q_{lm} \frac{d^l c_{lm}}{dr^l} \Big|_{r = 0}.
\end{equation}
This result allows us to easily extract the delta-function structure of the far field of an ideal multipole.  We see that an order-$l$ multipole has an order-$l$ \emph{derivative} of a delta function at the origin.

For example, for an order $l=0$ multipole (a monopole) with charge $q$, we only need to know the value of $f(\bm{x})$ at the origin: $f(\bm{x}) = c_{00}(r) Y^*_{00}(\Omega) + \dots = (4 \pi)^{-1/2} c_{00}(0) + o(r)$.  With our choice of normalization conventions \cite{Jackson}, $q_{00} = \sqrt{4 \pi}\, q$ so
\begin{align*}
\int_V d^3x\, \left[ f\, \bm{\nabla} \cdot \bm{E}_\text{ff}^{(00)} \right] &= q_{00}\, c_{00}(0) = 4 \pi q\, f(\bm{0}) \\
\bm{\nabla} \cdot \bm{E}_\text{ff}^{(00)}(\bm{x}) &= 4 \pi q\, \delta^3(\bm{x}) \\
\rho(\bm{x}) &= q\, \delta^3(\bm{x}).
\end{align*}

In the case of a dipole $\bm{p} \parallel \hat{z}$, we have \cite{Jackson} $q_{10} = \sqrt{4 \pi / 3}\, p$ and we need to keep the term
\begin{align*}
f(\bm{x}) &= c_{10}(r)\, Y^*_{10}(\Omega) + \dots \\
&= \sqrt{\frac{3}{4 \pi}}\, \cos \theta\, c_{10}(r) + \dots
\end{align*}
in expansion \eqref{f}.  The easiest way to proceed is to Taylor expand $c_{10}(r)$ (recalling that $c_{10}(0) = 0$) and then convert to Cartesian coordinates:
\begin{align*}
f(\bm{x}) &= \sqrt{\frac{3}{4 \pi}} \cos \theta\, c_{10}'(0)\, r + \dots \\
&= \sqrt{\frac{3}{4 \pi}} c_{10}'(0)\, z + \dots \\
\frac{\partial f}{\partial z} \Big|_{\bm{x} = 0} &= \sqrt{\frac{3}{4 \pi}} c_{10}'(0).
\end{align*}
Combining this with \eqref{master},
\begin{align}
\int_V d^3x\, \left[ f\, \bm{\nabla} \cdot \bm{E}_\text{ff}^{(10)} \right] &= \frac{8 \pi}{3} p\, \frac{\partial f}{\partial z} \Big|_{\bm{x} = 0} \nonumber \\
\bm{\nabla} \cdot \bm{E}_\text{ff}^{(10)}(\bm{x}) &= -\frac{8 \pi}{3} \bm{p} \cdot \bm{\nabla} \delta^3(\bm{x}). \label{dipoleDiv}
\end{align}

Even though $\bm{E}_\text{ff}^{10}(\bm{x})$ does not have a delta function at the origin (under spherical regularization), its divergence nevertheless has the derivative of a delta function at the origin!  We come to the surprising conclusion that by taking the divergence of a singular vector field, it is possible to directly produce the derivative of a delta function without ever ``passing through'' a delta function (either implicit or explicit) in the vector field.

This resolves the apparent paradox discussed in Section~\ref{Paradox}: we see that the mistaken step in our reasoning was in assuming that because $\bm{E}_\text{ff}$ does not contain a delta function at the origin, $\bm{\nabla} \cdot \bm{E}_\text{ff}$ does not contain the derivative of a delta function.  But in fact the ``far field'' and ``near field'' terms on the RHS of \eqref{DelDotE} \emph{both} have nonzero divergence at the origin and contribute to the charge density.  The two terms on the RHS of \eqref{DelDotE} add up to $\bm{\nabla} \cdot \bm{E}(\bm{x}) = -4 \pi \bm{p} \cdot \bm{\nabla} \delta^3(\bm{x})$.  Gauss's law then gives the correct answer $\rho(\bm{x}) = -\bm{p} \cdot \bm{\nabla} \delta^3(\bm{x})$, in agreement with \eqref{rho}.  (Ref.~\onlinecite{Namias} derives a result equivalent to \eqref{dipoleDiv} using a very different method that, as mentioned above, raises subtle mathematical issues and becomes increasingly complicated for higher multipole moments, whereas the general result \eqref{master} captures the contributions from all moments at once.)

Refs.~\onlinecite{Blinder} and \onlinecite{Griffiths} give the field of an ideal dipole ``on the understanding that the [far-field] term applies only to the region \emph{outside} an infinitesimal sphere about the point $r = 0$.''  But in fact, the divergence of the far-field term $\bm{E}_\text{ff}$ contributes two-thirds of the charge density at the origin, and this contribution is necessary for the consistency of the theory.  The far-field term is therefore important for capturing the physics arbitrarily close to the dipole, and the claim above is arguably an oversimplification.

If the arbitrary function $f(\bm{x})$ itself happens to obey the Laplace equation, then $c_{l'm'}(r) \propto r^{l'}$ for all $l' \geq 0$.  In this case, the surface term \eqref{surf} is actually independent of the radius $R$ of the region of integration, and the volume term \eqref{volumeInt} becomes zero, so only the surface term contributes to \eqref{intfDivF}.  Since the surface term does not go to zero at long distances, we must be careful to always retain it when integrating by parts, and this can pose some subtleties.  For example, when we considered the potential energy of a dipole in a uniform external field in \eqref{U}, we directly used the expression for the dipole charge density $\rho$.  If we instead use Gauss's law to express it in terms of the dipole field, we get
\begin{align*}
U &= \int d^3x\, \phi_\text{ext}(\bm{x}) \rho(\bm{x}) \\
&= \frac{1}{4 \pi} \oint d\bm{S} \cdot \left( \phi_\text{ext} \bm{E}_\text{ff} \right) + \frac{1}{4 \pi} \int d^3x\, \bm{E_\text{ext}} \cdot \bm{E} \\
&= \frac{1}{4 \pi} \oint d\bm{S} \cdot \left( \phi_\text{ext} \bm{E}_\text{ff} \right) \\
&\hspace{11pt}+ \frac{1}{4 \pi} \int d^3x\, E_\text{ext,i} \frac{3 n_i\, p_j n_j - p_i}{r^3} - \frac{1}{3} \bm{p} \cdot \bm{E}_\text{ext}(\bm{0}) \\
&= -\frac{2}{3} \bm{p} \cdot \bm{E}_\text{ext}(\bm{0}) + 0 - \frac{1}{3} \bm{p} \cdot \bm{E}_\text{ext}(\bm{0})
\end{align*}
from \eqref{surf} and \eqref{volumeInt}.  The surface term, which we usually drop in the second line, actually contributes two-thirds of the potential energy in this case, so we need to keep it in order to get the right answer.  (The surface and volume terms in the second line above contribute two-thirds and one-third of the total potential energy respectively, regardless of the regularization procedure, but the regularization procedure will determine how the volume contribution is distributed between its two terms.)

Finally, there is also a shortcut for calculating \eqref{dipoleDiv}, although making it rigorous is nontrivial.  Applying Gauss's law to \eqref{mixedPartialE} gives
\[
\rho(\bm{x}) = \frac{1}{4 \pi} p_j \partial_i \partial_i \partial_j \left( \frac{1}{r} \right) = p_j C_j(\bm{x}),
\]
where $C_j(\bm{x}) := 1 / (4 \pi)\, \partial_i \partial_i \partial_j \left( 1 / r \right)$.  \eqref{mixedPartials} then gives
\begin{equation} \label{C1}
C_j(\bm{x}) = \frac{1}{4 \pi} \partial_i \left( \frac{3 n_i n_j - \delta_{ij}}{r^3} \right) - \frac{1}{3} \partial_j \delta^3(\bm{x}).
\end{equation}
$C_j(\bm{x})$ is clearly an extremely pathological distribution, and we cannot simply assume that its mixed partial derivatives commute.  But \emph{distributional} derivatives always commute, so we can get the alternative expression
\begin{equation} \label{C2}
C_j(\bm{x}) = \frac{1}{4 \pi} \partial_j \left( \partial_i \partial_i \left( \frac{1}{r} \right) \right) = -\partial_j \delta^3(\bm{x}).
\end{equation}
(Evaluating the derivatives in this order roughly physically corresponds to first finding the charge distribution corresponding to an ideal monopole, then spatially differentiating that distribution to ``split'' the monopole into an ideal dipole.) Equating \eqref{C1} and \eqref{C2} gives
\[
\partial_i \left( \frac{3 n_i n_j - \delta_{ij}}{r^3} \right) = -\frac{8 \pi}{3} \partial_j \delta^3(\bm{x}),
\]
and contracting both sides with $p_j$ gives \eqref{dipoleDiv}.

\section{Conclusion \label{Conclusion}}

The delta-function structure at the origin of an ideal order-$l$ multipole field is significantly more complicated for $l \geq 1$ than for the monopole case $l = 0$: differentiating $\phi$ to get $\bm{E}$ produces a ``near-field'' order-$(l-1)$ \emph{derivative} of a delta function, then differentiating $\bm{E}$ to get $\rho$ produces a second, order-$l$ derivative of a delta function from the ``far-field'' term, and we need to keep track of both in order to get the correct charge distribution.  The near-field and far-field terms are therefore more subtly entwined than their names might suggest.  Fortunately, the single simple formula \eqref{master} captures the far-field contribution for any multipole moment.

\begin{acknowledgments}
The author would like to thank Mark Srednicki, Brayden Ware, Alex Rasmussen, and Dominic Else for helpful discussions, and Vladimir Hnizdo and Bruno Klajn for bringing Refs.~\onlinecite{Estrada, Namias} to the author's attention.
\end{acknowledgments}

\bibliography{Bib}
\bibliographystyle{apsrev4-1}

\end{document}